# Resonant tunneling in GaAs/$Al_YGa_{1-Y}As$ triple-barrier structures under uniform transverse magnetic field

Petrica I. Cristea

*University of Bucharest, Department of Physics, Bucharest-Mãgurele MG-11, ROMANIA*

*E-mail: pcristea@physicist.net*

**Abstract**. A transfer-matrix approach is used to simulate numerically the effect of a uniform transverse magnetic field on resonant tunneling in a symmetric GaAs/$Al_YGa_{1-Y}As$ triple-barrier resonant tunneling structure (TBRT) with Y=0.6, barrier regions of 30 A, and quantum well regions of 45 A. The resonant lines splitting (and hence the coupling energy vs. magnetic induction), for the ground and the first excited resonant doublet, shifts up in energy by increasing the magnetic induction. It is expected that, at very high fields, the interaction with confined phonons be enhanced, due to phonon coupling between electron states in the ground quasibound doublet. Each of branches in the resonant dispersion relations shows a parabolic behavior with the location of the cyclotron orbit center. At moderate magnetic fields, the time delay associated with the resonant tunneling is not considerably affected.

## 1. Introduction

Advanced crystal growth techniques, such as metal-organic vapour phase epitaxy (MOVPE) and molecular beam epitaxy (MBE), make it possible to obtain quantum wells and superlattices with reproducible properties at the atomic scale. By a further reduction in the dimensionality, new electronic properties are revealed in one-dimensional quantum wires [1] and zero-dimensional quantum dots (the ultimate quantum confinement structure) [2,3]. In the past few years much attention has been focused on the physics of quantum mechanical resonant tunneling through GaAs/AlGaAs double and multibarrier systems. In absence of impurities, at low temperature, the phase coherence length is larger than the sample size and the wave nature of the electrons needs to be taken explicitly into account [4]. The coupling between adjacent wells results in transport through miniband states. If a magnetic field $\boldsymbol{B}$ is applied, the wave vector $\boldsymbol{k}$ moves in a direction perpendicular both to $\boldsymbol{B}$ and to the gradient of the energy surface, which means it follows a contour of constant energy about the field direction. With magnetic field, applied in the same direction as the confining electric field, a magnetic quantization in the layer plane occurs, and the two-dimensional (2D) electron gas collapses to a discrete set of cyclotronic Landau orbits of zero degrees of freedom (0D). In such a case the Hamiltonian separates into an electric part giving rise to subbands, and a magnetic part leading to Landau levels. This case has been extensively studied concerning the quantized Hall effect [5-7] and Shubnikov-de Haas oscillations [8]. If any other configuration is used, this separation is not possible The effect of a transverse magnetic field on the tunneling through a barrier separating two semiconductors, Snell *et al* [9], or two superlattices, Davies *et al* [10], also has attracted great attention and has stimulated many experimental studies of magnetotransport. When the magnetic induction ***B*** is applied perpendicular to the direction of the tunneling current density ***j*** (z-axis in fig.1), even without impurities and any scattering mechanisms, the momentum along the interfaces is no longer conserved, but modified by the Lorentz force. This is a semiclassical picture and such an effect can be thought of as modifying the form of the barrier potential in absence of the field. A remarkable situation occurs if the magnetic length becomes comparable to the quantum-well width, because the confining electric field and the magnetic field contribute almost the same weight to the energy levels of electrons [11]. By taking the advantage that the harmonic oscillator is one of the exactly solvable problems, Lee *et al* have calculated the exact eigenenergy spectrum of an electron in a quantum well within an in-plane magnetic field.

In this letter we study how a uniform transverse magnetic field influences the tunneling through GaAs/AlGaAs triple-barrier resonant tunneling structure (TBRT). This material is organized as follows: Section 2 gives details of geometry, composition and band-diagram for the TBRT device. Section 3 outlines the theoretical approach. In Section 4 we present numerical results showing the influence of the transverse magnetic field on the transmission probability and coupling energy. The electron resonant energy dispersion relation is presented in Section 5. Finally, in Section 6, we present numerical results for the coherent scattering phase shift and the time delay associated with the passage of a particle through a TBRT device.

## 2. TBRT structures

The conduction-band energy diagram for a TBRT structure is shown Fig.1(a,b). By simply eliminating the coupling barrier, a TBRT structure transforms into a double-barrier (DBRT) one. These structures consist of two heavily doped $n^+$ GaAs layers emitter and collector ($\approx 2\text{x}10^{17}$ $\text{cm}^{-3}$), undoped AlGaAs barriers (30A, $Y$=0.6) and the undoped GaAs quantum well (QW) regions (45A). The conduction-band offset $\Delta E_C$, effective mass $m^*(z)$ and dielectric constant $\varepsilon(z)$ in each region of the DBRT or TBRT structure are determined as function of the aluminum concentration $Y(z)$, by the following approximations [12]:

$$\Delta E_C(z) = 0.75Y(z) \text{ eV for } 0 \le Y(z) \le 0.45 \quad (1a)$$

$$\Delta E_C(z) = 0.75Y(z) + 0.69[Y(z)-0.45]^2 \text{ eV for } 0.45 < Y(z) \le 1 \quad (1b)$$

$$m^*(z)/m_O = 0.067 + 0.083Y(z) \text{ for } 0 \le Y(z) \le 1 \quad (1c)$$

$$\varepsilon(z)/\varepsilon_O = 13.1 - 3.0Y(z) \quad \text{for} \quad 0 \le Y(z) \le 1 \qquad (1d)$$

### *2.1 Two coupled quantum wells, without interaction with the continuum*

When the thickness of the top and bottom barriers tends to infinite, an isolated system with two coupled quantum wells (of width $d_W$) is obtained. We introduce the notations: $\theta_b = k_b d_w$, $i\theta_w = k_w d_w$, $c_b = d\sqrt{2m_b^* V_0}/\hbar$, $c_w = d\sqrt{2m_w^* V_0}/\hbar$, where $k$ is the $z$-component of the wave vector, $d$ is the thickness, and *'b'*, *'w'* stand for the barrier and quantum well regions, respectively. With the transfer matrix method, and a boundary condition that conserves carrier current [the continuity of the anvelope wavefunction $\varphi(z)$, and its first derivative divided by $m^*(z)$, $\varphi(z)'/m^*$ ] we obtain a system of equations for the eigenenergies of the ground and the excited bounded doublets:

$$\begin{cases} 2\dfrac{m_b^*}{m_w^*}\theta_b\theta_w \cos\theta_w + \left[\theta_b^2 - \left(\dfrac{m_b^*}{m_w^*}\theta_w\right)^2\right]\sin\theta_w = \pm\left[\theta_b^2 + \left(\dfrac{m_b^*}{m_w^*}\theta_w\right)^2\right]exp(-a\theta_b/d_w)\sin\theta_w \\ \dfrac{\theta_b^2}{c_b^2} + \dfrac{\theta_w^2}{c_w^2} = 1 \end{cases} \qquad (2)$$

This system has been solved vs the width of the coupling barrier for a TBRT structure with $d_W$=60 A, $V_O$=0.536 eV (corresponding to an aluminum concentration of $Y$=0.67). The eigenenergies vs coupling are presented in the Table I.

**TABLE I**

| The width of the coupling barrier<br>*a (A)* | Eigenenergies (eV)<br>$E_{0,1}$<br>$E_{0,2}$ | <br>$E_{1,2}$<br>$E_{1,2}$ |
|---|---|---|
| 5 | 0.04969<br>0.09197 | 0.25151<br>0.36799 |
| 10 | 0.06110<br>0.08416 | 0.26522<br>0.33918 |
| 20 | 0.06982<br>0.07665 | 0.28154<br>0.31136 |
| 40 | 0.07297<br>0.07358 | 0.29236<br>0.29741 |
| 60 | 0.07325<br>0.07331 | 0.29439<br>0.29525 |
| ∞ | 0.07328 | 0.29482 |

## 3. Theory

We assume a constant magnetic field in the x-direction ***B****=(B,0,0)*. This can be represented by a vector potential in the gauge ***A****=(0,-Bz,0)*. Then the following Hamiltonian describes a spinless particle in a layer of the structure, with an effective mass *m*(z)* and charge *-e,* subject to a constant and uniform transverse magnetic field ***B***:

$$H = -\frac{\hbar^2}{2m(z)^*}\Delta - ie\frac{\hbar}{m(z)^*}\mathbf{A}\cdot\bar{\nabla} + \frac{e^2\mathbf{A}^2}{2m(z)^*} + V(z) \tag{3}$$

where *V(z)* is the potential-energy seen by a single electron, which includes effects of both conduction-band discontinuities at GaAs/$Al_YGa_{1-Y}$As interfaces and external applied bias. The incoherent electron scattering, space-charge effects, many-electron effects and phonon-assisted tunneling are neglected. However, it should be noted that the interaction with phonons may be responsible for the satellite peaks in DBRT current, at voltages just above the resonant peak [13,14]. The stationary Schrödinger equation corresponding to the Hamiltonian (3) is therefore $H\Psi(\mathbf{r}) = E\Psi(\mathbf{r})$. Notice that the choice of vector potential is not unique for the given magnetic field. With a different one the solutions would then look very different while the physics must remain the same. It is only with our choice of gauge, that the solutions have translation symmetry in the *x* and *y* directions. Therefore the wave function $\Psi(\mathbf{r})$ can be written as a product of a plane wave with an anvelope wavefunction *φ(z)* describing the motion of the tunneling electrons along the *z*-direction of the structure $\Psi(\mathbf{r}) = \varphi(z)exp\left[i(K_x x + K_y y)\right]$. By substituting this wave function into the stationary Schrödinger's equation we obtain a basically one-dimensional Schrödinger equation for the anvelope wavefunction *φ(z)*:

$$\frac{d^2\varphi}{dz^2} + \frac{2m^*}{\hbar^2}\left\{\left(E - \frac{\hbar^2 K_x^2}{2m^*}\right) - \left[V(z) + \frac{e^2B^2}{2m^*}(z - z_o)^2\right]\right\}\varphi = 0 \tag{4}$$

We note that the influence of the magnetic field is included by changing the ordinary superlattice potential and external applied bias effects *V(z),* with an effective one dimensional potential $V_{eff}$ (see fig.1,a,b):

$$V_{eff} = V(z) + \frac{1}{2}m^*(z)\omega_c^2(z)(z - z_o)^2 \tag{5}$$

where $z_o = l_B^2 K_{yo}$, with $l_B = (\hbar / eB)^{1/2}$ the magnetic length, and $\omega_c(z)$ stands for the cyclotron frequency associated with a *z*-dependent effective mass *m*(z)*. The parameter $z_o$ gives the center of the cyclotron orbit. Due to ionized impurity scattering, the coherence of the Landau motion is destroyed in emitter and collector regions. Hence it is expected that the effect of the magnetic field in these regions be small enough to set for the magnetic vector potential $\boldsymbol{A}_{emitter} = \boldsymbol{A}_{collector} \cong 0$. We used a transfer-matrix ( TM ) method [15 , 16] to solve this one-dimensional Schrödinger equation and to calculate the transmission probability, resonant linewidths, the transit time and the anvelope wavefunctions $\varphi(z)$ for the motion along z-axis, under or without applied bias $V_a$ and magnetic field $\boldsymbol{B}$, perpendicular to *z*-axis. This method has been widely used due to its simplicity related with the use of only 2x2 matrices and with the possibility of studying superlattices formed by *any sequences of layers*. We discretize the barrier and the quantum-well (QW) regions into a finite number of steps, so that, at any step, a flat-band potential approximation can be used. Ando and Itoh [15] have shown that as the number of steps increases and the new step-like potential will be closer and closer to $V_{eff}(z)$, the solution rapidly converges to a single result. Therefore, at any step, the solution to the Schrodinger equation (4) is given as a superposition of waves $\varphi(z) = a \cdot exp(K_z z) + b \cdot exp(-K_z z)$, with the *z* component of the wave vector $K_z$ given by:

$$K_z(z) = \sqrt{\frac{2m(z)^*}{\hbar^2}\left\{\left[V(z) + \frac{e^2B^2}{2m(z)^*}(z - z_o)^2\right] - \left(E - \frac{m_o^*}{m(z)^*}E_{x,o}\right)\right\}} \tag{6}$$

where $m^*_o$ and $E_{x,o}$ are the effective mass and the kinetic energy along *x* axis, in the emitter region. Using a boundary condition that conserves carrier current [the continuity of the anvelope wavefunction $\varphi(z)$, and its first derivative divided by $m^*(z)$, $\varphi(z)'/m^*$ ] , the coefficients $a_j$ and $b_j$ of region *j* are joined to those of region $j+1$ by a 2x2 transfer matrix $\boldsymbol{T}_{j,j+1}$ of determinant 1:

$$\begin{pmatrix} a_{j+1} \\ b_{j+1} \end{pmatrix} = \mathbf{T}_{j,j+1} \begin{pmatrix} a_j \\ b_j \end{pmatrix} \tag{7}$$

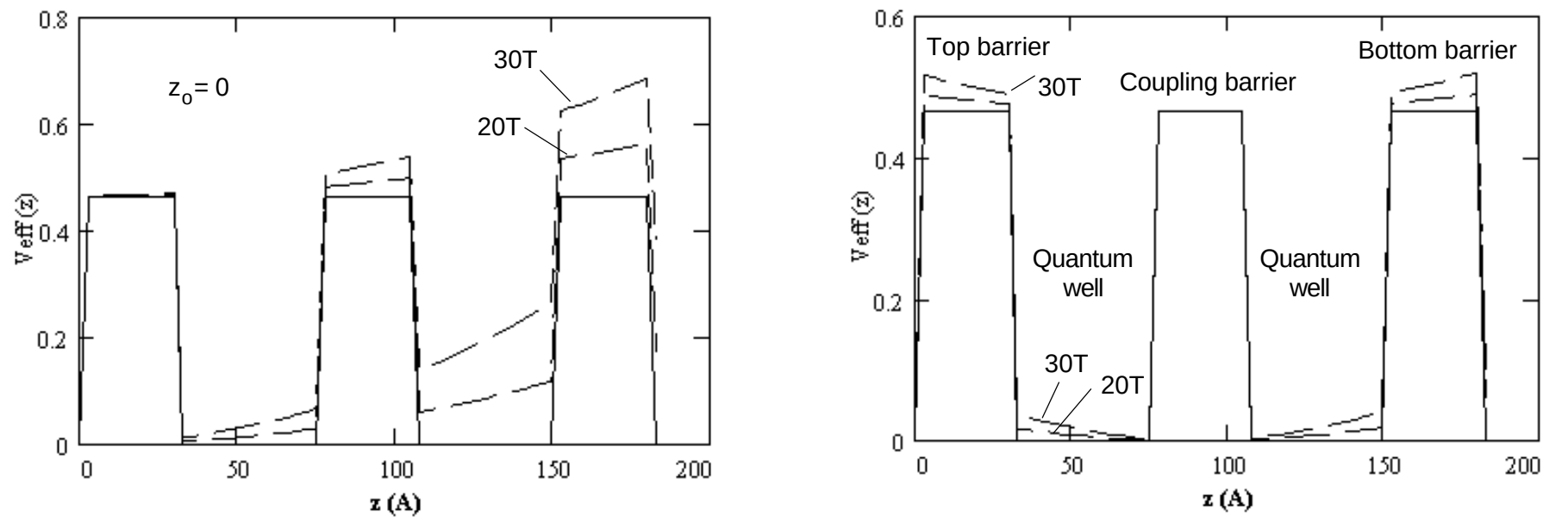

Fig.1. The effective one-dimensional potential *V(z)* (in eV) as a function of transverse magnetic field *B* in a symmetric triple barrier resonant structure with $L_{B1}=L_{B2}=L_{B3}$=30A, $Y_{B1}=Y_{B2}=Y_{B3}$=0.6, $L_{QW1}=L_{QW2}$=45A, for $z_o$=0 (a) and $z_o=L/2$ (b) The origin for $z_o$ lies at the front of the top barrier and *L* stands for the total thickness of the TBRT structure.

## 4. Transmission probability and coupling energy vs. magnetic induction

We have plotted in figures 2(a) and 2(b) the effect of an increasing transverse magnetic field **B** on the transmission probability, for the first and the second resonant doublet, respectively. The TBRT structure is the same as in Fig.1, with zero applied bias and $z_o=0$. In this case, for $B\neq 0$, the effective potential is no longer symmetric (see fig.1a) and therefore the transmission probability no longer achieves a peak value of 1. Sharp peaks occur in the transmission for the resonant energies $E_{01}$, $E_{02}$ (fig.2a) and $E_{11}$, $E_{12}$ (fig.2b). These are the energies corresponding to the ground quasibound doublet and the first excited one. Increasing the magnetic field the transmission peaks shift higher in energy. This is reasonable because the magnetic contribution to $V_{eff}$ is, somehow, equivalent to a reverse bias applied on the structure (see also fig.1a; numerical results prove that even for $z_o=L/2$ this shift to higher energies also holds, but the magnetic contribution is no longer equivalent to an applied bias, see fig.1b). Figures 3(a), 3(b) show a plot of the resonant lines versus magnetic induction $B$, for the first and second doublet, respectively. The results presented show that the splitting between the resonant lines (and hence the coupling energy vs $B$), for each resonant doublet, also shifts up in energy by increasing the magnetic induction. It should be noted that the coupling energy for the ground quasibound doublet is much more affected by the magnetic field than the excited one. These doublets are generated by splitting the symmetric ground states and the antisymmetric excited ones of an isolated quantum well into a symmetric-antisymmetric pair. The symmetric states have, as expected, lower energy. Such a splitting is caused by the coupling between the wells (this means that the degeneracy of each level is removed due to a coupling barrier of finite thickness). Also, each resonant level has a finite width induced by the coupling with the left (emitter) and right (collector) justified states in the continuum spectrum. Moreover, in numerical simulations, a Gaussian broadening is generally included to simulate the presence of disorder and nonvanishing temperatures.

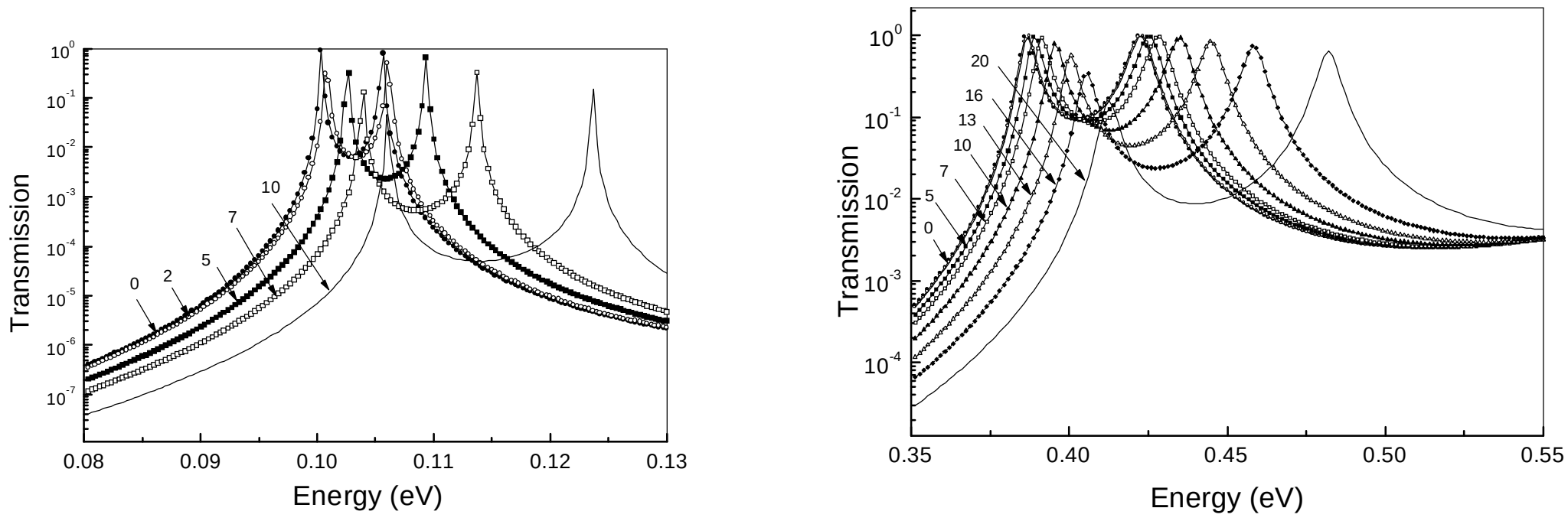

a) b)

Fig.2. The effect of an increasing transverse magnetic field on the transmission probability, for the first (a), and the second (b) doublet. The TBRT device is the same as in fig.1, with no applied bias and $z_o=0$. Numbered arrows indicate the magnetic induction in units of Tesla.

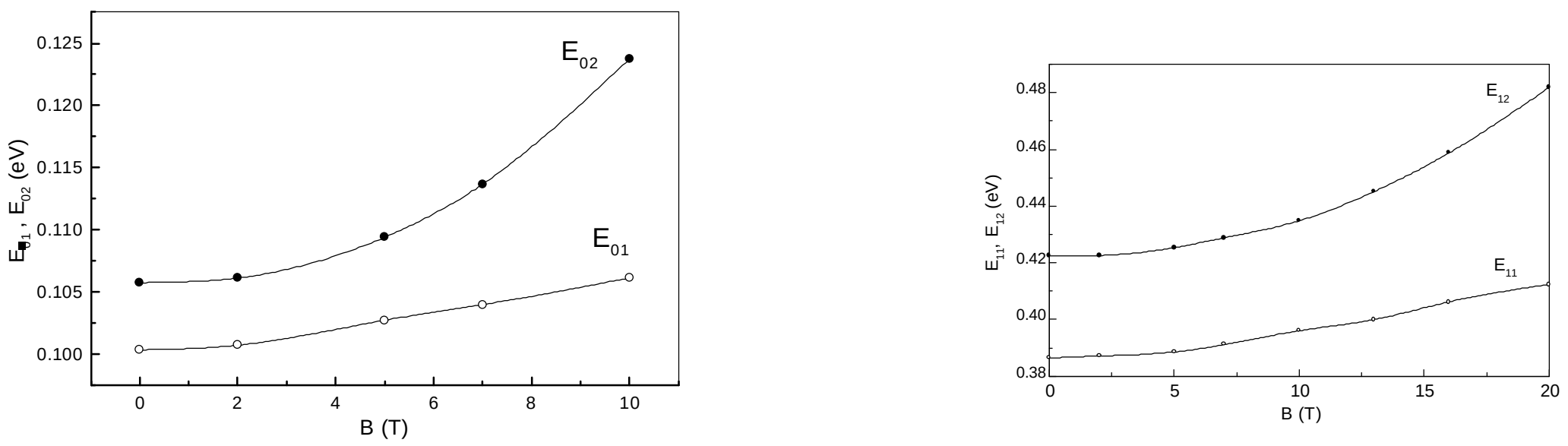

a) b)

Fig.3. Resonant lines versus magnetic induction, for the ground doublet (a) and the excited doublet (b) ($z_o=0$, without applied bias). The structure is the same as in fig.2

The presented results neglect this later contribution. Note that, by increasing the thicknesses of the top and bottom barriers, the results are rapidly converging to those describing an isolated system with two coupled quantum-wells or an isolated quantum well (when the coupling barrier is removed). Such a numerical trick avoids some mathematical difficulties encountered by the methods using some analytical solutions of equation (4).

## 5. Electron resonant energy dispersion relation

We have plotted in figure 4 the calculated resonant energy dispersion relation of the doublets ($E$ as a function of $z_o$) in a TBRT structure (with the same parameters as in fig.1) for $B$=5T, without applied bias. In this figure $z_o$ is measured with respect to the center of the coupling barrier, in dimensionless units of $z_o/L$. Each branch of the dispersion relations shows a parabolic behavior $E=e_1+e_2(z_o/L)^2$, the coefficients $e_{1,2}$ being listed (for $B$=5T) in Table II. These coefficients generally depend on the magnetic field intensity and structure parameters, but the parabolic behavior is conserved even when the magnetic field considerably increases. Without magnetic field the dispersion relations for resonant energies are almost flat. For comparison, the dotted lines represent the dispersion relation of the ground and the first excited resonance in a DBRT structure, the QW and barrier regions having the same parameters as in the TBRT structure. We note that for such a DBRT structure the energy dispersion relations are less affected by the magnetic field.

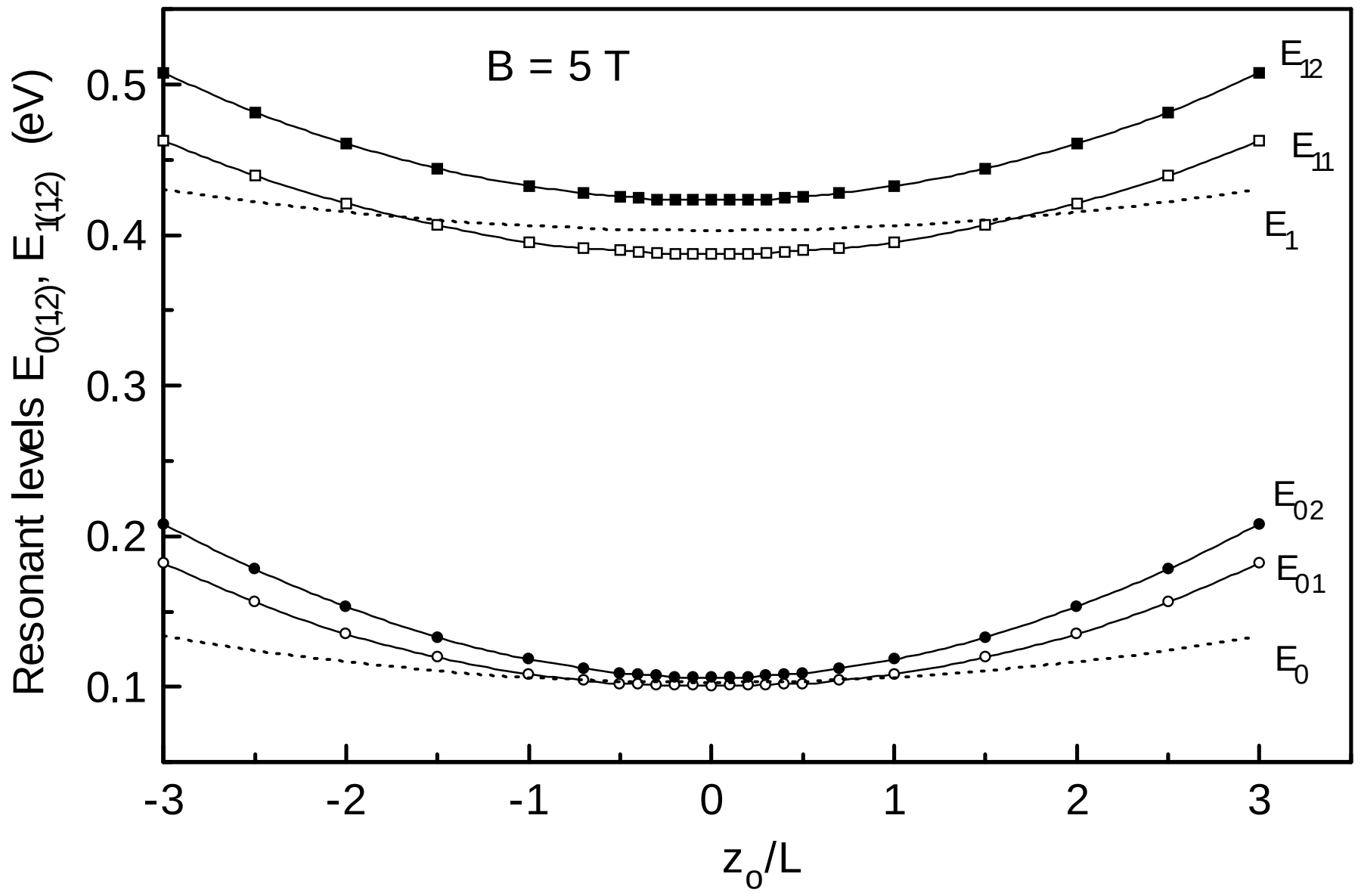

Fig.4. Resonant energy dispersion relation for tunneling electrons in a TBRT structure (the parameters are the same as in fig.1), $B$=5T, without applied bias. Open circles $E_{01}$, full circles $E_{02}$, open squares $E_{11}$, full squares $E_{12}$. The dotted lines correspond to the dispersion relation in a DBRT structure with QW and barrier regions having the same parameters as in the TBRT structure. $z_o$ is measured with respect to the centre of the coupling barrier.

**TABLE II**

| **Branch** | $e_1$ (eV) | $e_2$ (eV) |
|---|---|---|
| $E_{01}$ | 0.1001775 | 0.00900 |
| $E_{02}$ | 0.1066523 | 0.0113805 |
| $E_{11}$ | 0.3874347 | 0.008351 |
| $E_{12}$ | 0.42318 | 0.0094 |

## 6. Scattering phase shift

The existence of a time associated with the passage of a particle through a potential barrier was first suggested sixty years ago by MacColl [17]. Although is no clear consensus in the field, such a quantum time is now well accepted and some authors claim that this time has been indeed measured experimentally. However, the existence of a simple expression for this quantity as well the exact nature and meaning of that expression still remain open questions. The proposed expressions fall into three main classes [18]: i) the authors argue that expressions based on following a feature of a wave packet through the barrier have little physical significance; ii) a second class tries to identify a set of classical paths associated with the quantum mechanical motion and then tries to average over these; iii) the third class invokes a physical clock involving degrees of freedom besides that involved in tunneling.

An other important property of coherent transport through resonant devices is the scattering phase shift $\alpha = a\tan\{\mathrm{Im}[\varphi(L^+)]/\mathrm{Re}[\varphi(L^+)]\}$. Figure 5 shows the phase shift $\alpha$ as a function of incident electron energy in an unbiased TBRT structure (with the parameters as in Fig.1), ***B***=0, and in the energy range associated with the resonance of the first quasibound doublet. It should be noted that for most energy range the phase shift shows no relevant dependence on energy. However it gradually increases and around resonances it rapidly shift through $\pi$ radians. To avoid the conceptual difficulties discussed in the introduction of this section, we adopt a simple model and the phase shift α is related to time delay τ(E), experienced by a wave packet, through the approximation [19]:

$$\tau(E) \approx \hbar \frac{d\alpha}{dE} \tag{8}$$

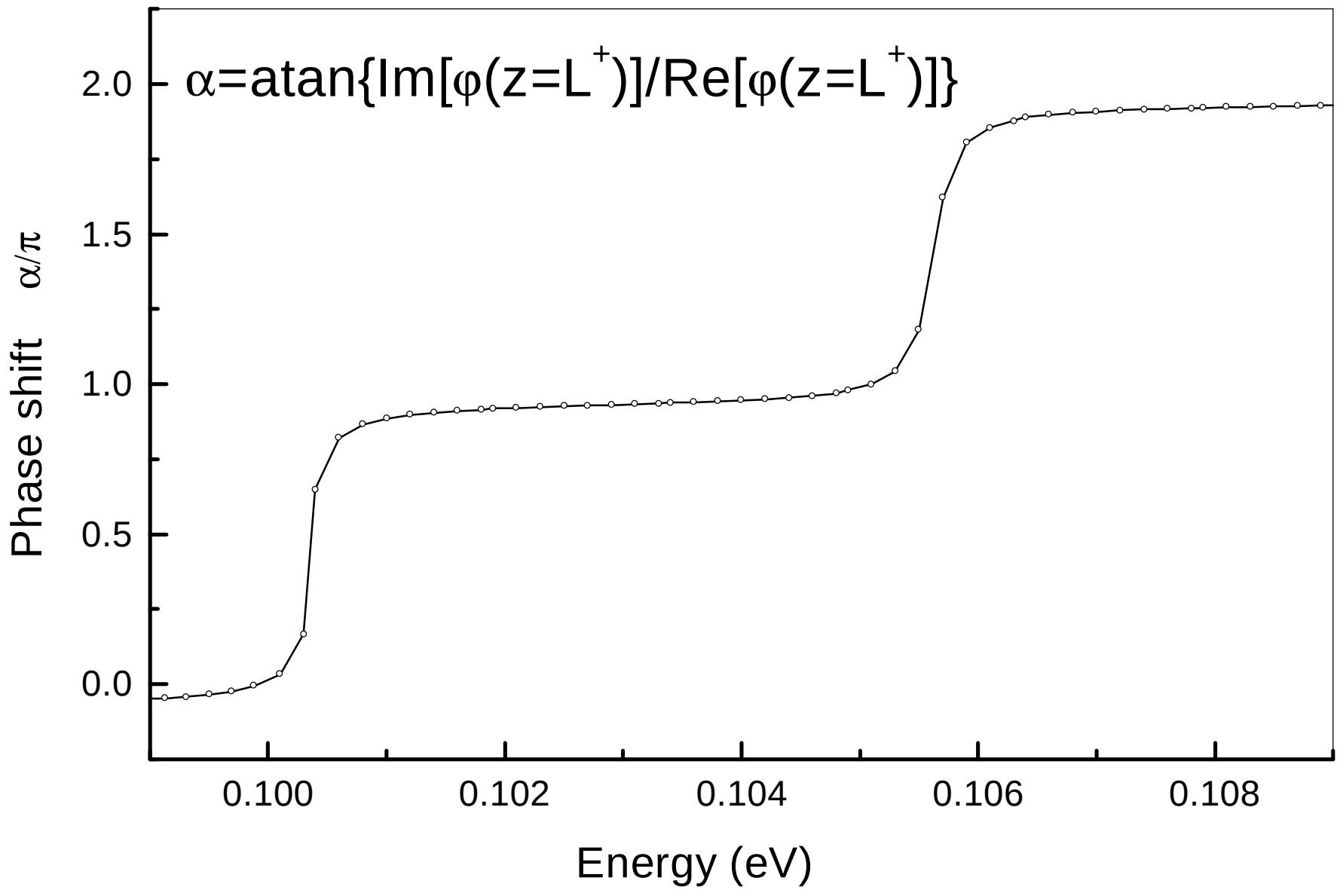

Fig.5. The phase shift $\alpha$, as a function of incident electron energy, in an unbiased TBRT structure (with the parameters as in fig.1), ***B***=0, and in the energy range associated with the resonance of the first quasibound doublet.

Therefore, for most energies, the time delay is rather small. However, near resonances the wave packet spends significantly more time in the TBRT structure. Figure 6 shows a numerical calculation of the time delay versus incident electron energy, for the device in fig 1, evaluated from equation (8), with no magnetic field and for *B*=7 T, near the resonance of the first quasibound doublet. Leaving aside the peaks shift toward higher energies, introduced by the presence of the magnetic field, the curves are remarkably similar. All these peaks show a lorentzian behavior:

$$\tau(E) = \frac{S}{\pi} \frac{\Delta E/2}{(\Delta E/2)^2 + (E - E_O)^2} \tag{9}$$

where $S \cong \hbar\pi$ (from eq.8) is the peak area, *ΔE* (≅$2\text{x}10^{-4}$ eV) is peak width, and $E_O$ is the resonance energy. Note that the peak width is related to resonance delay time $\tau(E_O)$ through a simple equation $\tau(E_O)\Delta E/2 = \hbar$. This means that the "resonance life-time" considerable increases in systems with wide barrier regions. Also, because in the resonance region the time delay $\bar{\tau}$ is obtained by averaging *τ(E)* over the line shape:

$$\bar{\tau} = \frac{1}{\gamma \Delta E} \int \tau(E) dE \tag{10}$$

where $\gamma$ is a dimensionless scaling parameter, one obtains $\bar{\tau} = \pi\tau(E_o)/2\gamma$. It should be noted that, by properly setting the scaling parameter, the equality $\int \tau(E)dE \cong \hbar\pi$ must hold (eq. 8).

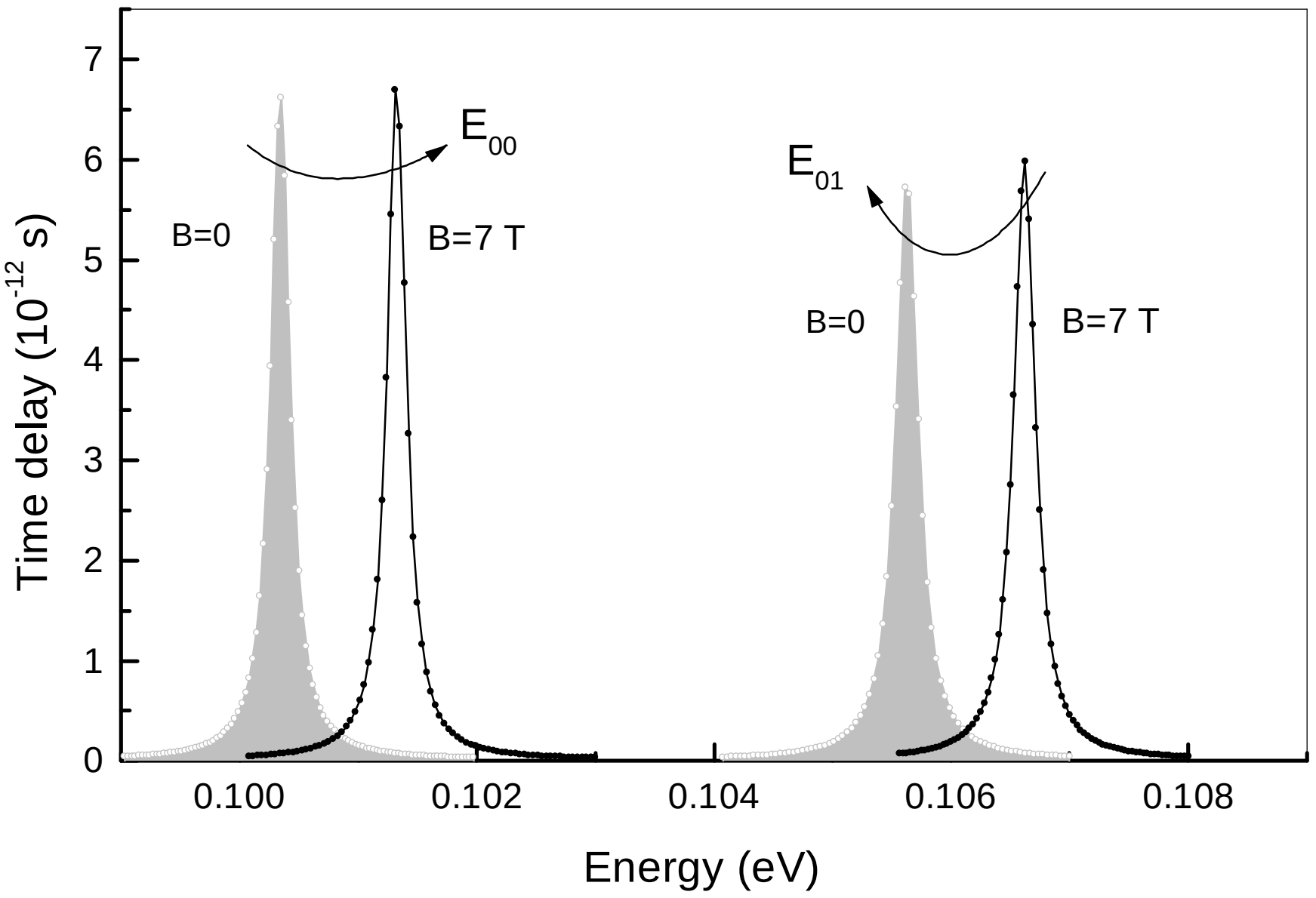

Fig.6. The time delay versus incident electron energy, for the device in fig 1, evaluated from equation (8), with no magnetic field (shaded peaks), and for *B*=7 T, near the resonance of the first quasibound doublet.

## 7. Conclusions

In this letter we have numerically examined the effect of a uniform transverse magnetic field on resonant tunneling in GaAs/$Al_YGa_{1-Y}$As triple-barrier structures. We have shown that under increasing the magnetic field the transmission peaks shift up in energy. The splitting between the resonant lines (and hence the coupling energy vs *B*), for each resonant doublet, also shifts up in energy. Is thus expected that, at very high *B* fields, the interaction with confined phonons will play a significant role due to the phonon coupling between electron states in the ground quasibound doublet. Each branch of the dispersion relations shows a parabolic behavior with the location of the center of the cyclotron orbit. At moderate magnetic fields, the resonance life time is not very much affected. For $Al_YGa_{1-Y}$As layers with $Y$>0.45 (we used throughout this work $Y$=0.6) the *X*-point minima of the conduction band have lower energy than the Γ-point minimum. Therefore, the tunneling electrons tend to scatter from the Γ-point during tunneling through the barrier regions. However, the Γ-to-X conversion becomes significant only for relatively thick barriers (> 50A) [16]. Because we used in simulations a TBRT structure with narrow barriers (30 A) this effect can safely be neglected. For the excited resonance doublet the effects of band nonparabolicity may become important. One way to overcome such a difficulty is to use an energy-dependent effective mass that is easily incorporated into the transfer-matrix formalism.